\documentclass[sigconf]{acmart}
\usepackage{array}
\usepackage{booktabs}
\usepackage{ragged2e}
\usepackage{multicol}
\usepackage{enumitem}
\usepackage{hyperref}
\AtBeginDocument{%
  }

\setcopyright{none}
\acmDOI{}
\acmISBN{}

\acmConference[LIMITS '25]{11th Workshop on Computing within Limits}{June 26--27, 2025}{Online}

\begin{document}

\title{A Protocol to Address Ecological Redirection for Digital Practices in Organizations}

\author{Valentin Girard}

\affiliation{\institution{Univ. Grenoble Alpes, Inria, CNRS, Grenoble INP , LIG ; \\
Univ. Grenoble Alpes, CNRS, Grenoble INP, G-SCOP,}
\country{38000 Grenoble, France}
}
\email{valentin.girard2@univ-grenoble-alpes.fr}

\author{Antoine Martin}
\affiliation{\institution{Independent researcher}\country{France}
}
\email{antoine.martin@sentier-ergonomie.fr}

\author{Maud Rio}
\affiliation{\institution{Univ. Grenoble Alpes, CNRS, Grenoble INP, G-SCOP,}\country{France, 38000 Grenoble}
}
\email{maud.rio@g-scop.eu}

\author{Romain Couillet}
\affiliation{\institution{Univ. Grenoble Alpes, Inria, CNRS, Grenoble INP , LIG,}\country{France, 38000 Grenoble}
}
\email{romain.couillet@univ-grenoble-alpes.fr}

\renewcommand{\shortauthors}{Girard et al.}

\begin{abstract}
The digitalization of societies raises questions about its sustainability and the socio-technical impacts it generates. Ecological redirection applied to organizations is a field of research aiming for achieving sustainability as a direction, rather than for technical means. Arbitration and renunciation to some digital usage and technologies are investigated. Ecological redirection is, however, not yet addressing concrete methodologies for its implementation in organizations. This paper therefore proposes a protocol to support stakeholders in the ecological redirection of their digital practices. This protocol is based on mapping attachments to digital tools through a multi-disciplinary survey. It then proposes increasing stakeholders' knowledge and skills to prepare a debate on the arbitration of renunciations, and finally, to operationalize the closure/transformation of targeted digital practices. This protocol will be tested in real conditions in different contexts. An empirical study is proposed to measure 1) the fluidity with which participants carry out the protocol, 2) the effectiveness of the protocol in terms of the redirection objective, 3) the socio-technical barriers to the redirection process. The paper concludes on the potential benefits for organizations to better understand both the barriers related to its ecological redirection and the transformative aim of such protocols. This will help them trigger large and radical policies towards a desirable and sustainable society.
\end{abstract}


\maketitle

\section{Introduction}
In response to the transgression of planetary boundaries on a global scale \cite{richardson_earth_2023}, various actors within the techno-industrial society are advocating for a digital transition (e.g. \cite{joint_european_2022}). This solution is often proposed in pairs with the energy transition \cite{garin_que_2023}. However, according to a report by the think-tank The Shift Project, ‘The digital transition as it is currently being implemented contributes more to climate disruption than it helps to prevent it’, notably because of indirect effects such as the rebound effect, induction, moral compensation and so on \cite{roussilhe_effets_2022}. The report further notes that ‘[these] global systemic effects [...] remain highly uncertain, despite often being considered positive a priori’\cite{ferreboeuf_lean_2018}.

Indeed, the share of emissions from the digital sector -- which encompasses the entire life cycle of devices, network infrastructures, and data centers -- is estimated to account for between 2.1\% and 3.9\% of total global greenhouse gas emissions \cite{freitag_real_2021}, with an annual growth rate of 8\% in 2018 \cite{ferreboeuf_lean_2018}. Besides these emissions, the sector requires a significant amount of metal resources, severely impacting the health of the ecosystems and the communities most exposed to its industry \cite{izoard_ruee_2024}. The entire infrastructure manufacturing process is highly energy-intensive and demands large quantities of drinking water \cite{bordage_sobriete_2019}. Furthermore, this industry, which is based on short-lived devices, is responsible for several tens of millions of tons of electronic waste each year: 62 million tons in 2022\footnote{With major regional disparities : 17.6 kg per capita per year in Europe, 14.1 in America, 6.4 in Asia and 2.5 in Africa.}, of which only 22\% were documented as formally collected for recycling in an environmentally sound manner \cite{balde_global_2024}.

Beyond the environmental and social impacts generated by the digital sector, studies suggest that this sector is condemned in the long term \cite{tomlinson_information_2017}. This vulnerability can be explained by the destruction of ecosystems essential to its existence, the complexity of the value chains underlying the infrastructure, the societal instability it generates, and the limited availability of mineral and energy resources \cite{girard_computing_2024}. Despite these findings, the expansion of digitalization is rarely questioned and continues to progress at an exponential rate \footnote{ +8\% annual increase in greenhouse gas emissions, +9\% annual increase in energy consumption, +40\% annual increase in data storage, +25\% increase in data flow \cite{ferreboeuf_lean_2018}}.

Given the urgency and the insufficient impact of current transition policies, the framework of ecological redirection proposes a change in perspective: rather than considering that we are facing an ecological crisis that requires technical solutions, it suggests that we are dealing with a new socio-climatic regime (the Anthropocene\footnote{This controversial concept developed in 2000 by J Crutzen \cite{steffen_anthropocene_2011}, highlights the significant impact of the human species on geology and ecosystems on the scale of the Earth's history \cite{steffen_trajectory_2015}.}), which raises primarily a question of direction\footnote{The concept of direction can be compared to Donella Meadows' third leverage point: the goals of the system \cite{meadows_leverage_1999}.}. This new regime necessitates both renunciations and closures of certain activities to reorganize our conditions of subsistence. Of course, not all activities should be closed, and stakeholders should be able to suggest certain arbitrations in the process of renunciation, based on criteria of (un)sustainability and vulnerability.

The field of ecological redirection is particularly developed in France\footnote{Open and lead at first by Alexandre Monnin, Emmanuel Bonnet and Diego Landivar after their original publication Héritage et Fermeture \cite{bonnet_heritage_2021}, inspired by theories of Tony Fry and Bruno Latour.}. Its approach is currently mainly theoretical and requires further specification, especially within the context of the digital sector: ‘[…] This situation, experienced by generations endogenous to the digital age, dramatizes the questions we can ask about the conditions for achieving sobriety, or even implementing active processes of detachment or renunciation. This question seems to us to open a considerable and urgently needed area of research’ \cite{allard_laurence_ecologies_2022}. Organizations, however concerned about the environmental impact of their digital tools, usually find it difficult to make a real ecological redirection due to a lack of efficient methodology.

This article thus proposes a protocol for the ecological redirection of digital tools and practices for organizations. It can be applied to any organization, whatever its size and scope, strategy, governance structure, incorporation structure or relationship to profit (these five dimensions of business have been theorized by Jennifer Hinton \cite{hinton_five_2021}). The protocol can be applied, for example, by publicly-traded shareholder corporations, producer cooperatives, worker cooperatives, dual purpose companies or not-for-profit structures.

Section 2 presents the framework of ecological redirection. Section 3 is a proposition for a protocol for the ecological redirection of digital tools and practices for organizations. Section 4 explores the application of this protocol to case studies, and outlines objectives for evaluating and improving the protocol. Section 5 concludes on a perspective regarding the challenges related to the ecological redirection of the digital sector.

\section{The Ecological Redirection Framework}

\subsection{Theoretical framework}
Ecological redirection is a political framework that has been developed in France, notably through the work of researchers Alexandre Monin, Emmanuel Bonnet, and Diego Landivar \cite{bonnet_heritage_2021}. This conceptual framework emerged as a response to the techno-solutionist ecological transition advocated by many actors in the 21st-century techno-industrial society. The ecological redirection framework distinguishes itself from techno-solutionism on four main controversial points \cite{marchand_introduction_2021}:

\begin{enumerate}
    \item Unlike the view that the ecological crisis is a problem that requires solutions, ecological redirection defends that we have entered a new Earth state -- the Anthropocene -- which is a complex phenomenon of destabilization of the Earth system which thus redefines the conditions of existence for humans and ecosystems.
    \item Rather than viewing the ecological crisis as a technical problem to be optimized, ecological redirection questions the direction of the societies and alerts us about the limitations of a techno-solutionist approach.
    \item Ecological redirection does not take for granted the technical possibility of achieving sufficient decoupling to return within planetary boundaries while maintaining the current societal model \cite{noauthor_decoupling_2019}. It argues that a redirection of our societal model is necessary to find ways of inhabiting an uncertain Earth system and to mitigate the degradation of Earth's habitability.
    \item The negative externalities of human activities are seen by ecological redirection as necessary conditions for maintaining these activities, thus viewing them as causes rather than consequences.
\end{enumerate}

In this way, ecological redirection shifts the focus from technical solutions to political direction. It advocates for a disconnection of society’s means of subsistence from the technosphere (i.e. the ‘realm of technology’, composed of ‘the totality of technological infrastructures’\cite{technosphere}), arguing that the latter destroys the conditions for human planetary habitability and condemns us in the long term, without any possible decoupling.

Ecological redirection is fully anchored in a political ecology paradigm, alongside frameworks such as degrowth \cite{hickel_less_2020}, decolonial ecology \cite{ferdinand_ecologie_2019}, ecofeminism \cite{mies_ecofeminism_2022}, and so on. It distinguishes itself from these other approaches, particularly through its focus on strategy and the operational framework.

\subsection{Strategic approach}

\subsubsection{Inheriting the Technosphere}\label{sec_technosphere}

As part of this goal to disconnect from the technosphere, ecological redirection emphasizes the importance of understanding individuals' attachments to it. These attachments are ‘what we hold on to and what holds us’ \cite{monnin_politiser_2023}. To be more precise, an attachment can be defined as a preference for (‘what we hold onto’) or constraint to (‘what holds us’) a mode of satisfying a need through a given artifact. It seems impossible to disconnect from the technosphere without considering the heritage it leaves behind, as we remain dependent on it in the short term. This heritage is both material (factories, waste, new climatic states, extractivist-based industry, infrastructure of all kinds, etc.) and immaterial (organizations, political models, economic models, shared culture, etc.). Therefore, it is crucial to understand the legacy that keeps the industrial society (including engineering systems) dependent on the technosphere to eventually disconnect from it.

\subsubsection{Politicizing Renunciation}

Furthermore, ecological redirection highlights the importance of renunciation in changing the direction of our societal model. These renunciations should not be imposed but chosen democratically. Participatory democracy thus becomes a central issue in ecological redirection, enabling collective decision-making on which attachments to prioritize for disconnection.

\subsubsection{Closing, Dismantling, Reducing}
Finally, renunciation entails a lot of different skills to support the detachment process in a practical way. A lot of fields are involved, such as engineering, logistics, sociology, law, economy, art, and so on. Detaching from the technosphere therefore requires various skills in closing, dismantling, disinvestment, and more. The art and science of planned closure and renunciation enable society to disconnect from the technosphere while ensuring it stops causing harm. This approach allows for staying within planetary boundaries without breaking individuals' subsistence networks (but rather reconfiguring them) and taking care of the negative commons\footnote{Negative commons refer to "negative" tangible or intangible "resources", such as waste, nuclear power plants, polluted soil or certain cultural heritages (the rights of a colonizer, etc.) \cite{bonnet_heritage_2021}} of the technosphere that continue to generate environmental and social harm even after abandonment (e.g. hazardous waste).

\subsection{Operational approach}
Ecological redirection gives an important place to practices of investigation and design.

Investigation is an essential practice for making the legacy of the technosphere concrete and enabling organizations to better understand the attachments that keep them dependent on the techno-industrial society. This can be conducted through social science practices (using qualitative, quantitative, or mixed approaches), but must also extend to other disciplinary fields to understand attachments in legal, psychological, material, and other terms. It allows for the creation of a mapping of attachments.

Design also occupies a central role in moving from theory to action. Ecological redirection emphasize the need to operationally implement ecological redirection protocols to avoid being confined to the realm of ideas. Design thus plays a key role in creating new ways of "doing without" or "doing with less" \cite{goulet_faire_2022}, transforming technical and organizational systems, or preventing the emergence of unsustainable practices.

In concrete terms, ecological redirection must propose protocols for democratic renunciation and closures, where investigation and design play central roles.

This article proposes such a protocol specified for digital technologies. It supports organizations in addressing the issue of digital de-escalation \cite{girard_computing_2024} through the lens of ecological redirection.

\section{Proposition : A Protocol for Digital Technologies}

This section proposes a protocol for the ecological redirection of digital tools and practices. Its aim is to operationalize the theoretical framework of ecological redirection for organizations, specifically focusing on the digital practices in their activities. Table \ref{table_summarize} summarizes the steps of this protocol. Additional resources needed to ensure the replicability of this protocol are available online\footnote{Currently available in French only. English-translated versions will be available soon: \url{https://lig-membres.imag.fr/girard15/protocole-de-redirection-ecologique/}}.

\subsection{Goal, Scope, and Prerequisites}
This protocol is intended for social organizations engaged in collective projects. These organizations can vary in nature, including companies, public institutions, associations, schools, markets, unions, cooperatives, clubs, and more. The protocol's objective is to guide the redirection of the digital tools and practices used by the organization in their shared goals, rather than addressing the actions of members separately. For example, if a primary school decides to follow this protocol, it would apply to all the school's activities -- such as pedagogy, relationships with parents, administration, cafeteria registration, etc -- but would not extend to the activities outside the scope of the school (children's access to digital technologies at home for example).

The organization must be able to propose a representative sample of its members, ideally selected randomly, to form a decision-making citizens' assembly (called “focus group” \cite{rio_exploring_2021}). This sample should be of a reasonable size (preferably fewer than 20-30 people) to facilitate workshops and ensure that all participants can engage in dialogue during debates. For smaller organizations, all members may participate, depending on the context. A formal involvement from focus group members to participate in all working sessions is necessary for the successful implementation of this protocol.

\subsection{Phase 1: Mapping of Attachments}\label{sec_mapping}
The first phase involves an ethnographic survey\footnote{Some methods to conduct such a survey are presented here \cite{lune_qualitative_2017}.} aimed at understanding how the organization's activities depend on their digital tools and practices. The aim of ethnographic research is to understand how a social group functions. It must involve a phase of reflexivity regarding its own preconceptions (our "a priori" about a particular situation), requiring the researcher to maintain an open mind and to frequently challenge their certainties. Although this survey does not occur at the very beginning of the sociological approach (as we have already defined the initial goals of the study), it must still go through a problematization phase that will help adapt the scope of the research to the studied organization. Thus, the ethnographic approach may differ greatly depending on the study field. For this phase, we recommend consulting methodologies in social sciences.

The investigation phase must ultimately allow for a mapping of attachments. Attachments are understood in accordance with the definition provided in Section \ref{sec_technosphere}, and may therefore range from regulatory constraints to psychological dimensions, including technical constraints, among others\footnote{We refer the reader to the notion of path dependency on this point \cite{arthur_increasing_1994}.}. Therefore, the survey must account for attachments through psychological and social mechanisms, as well as in material and organizational terms.

Although the investigative approach is, by definition, adaptive to the field studied and requires the researcher to build their operational approach, we propose a standard method from which they can draw inspiration. This standard approach suggests conducting three types of actions: 1) direct observation, 2) individual interviews, and 3) the implementation of a questionnaire. These actions will be conducted in parallel and must be adjusted in their implementation as the survey progresses, depending on the field conditions and the ongoing analysis, which may lead to new questions. The following paragraphs outline the proposed standard method, and these should not be considered as imperatives for conducting the investigation.

\subsubsection{Direct Observation}

This practice forms the foundation of the survey approach for any experimental field of our protocol. The aim of this practice is to directly observe (and potentially experiment) the actors’ practices in order to discover the general organizational mechanisms of the studied structure, begin to grasp the information system that support this organization, and identify initial points of reflection for the analysis of attachments. To achieve this, the researcher maintains a research journal, collecting field notes, analyses, and thought processes. This field journal will initially allow the researcher to problematize their investigation.

As the investigation progresses (after some observation sessions and interviews), it will be necessary to obtain more specific data on certain mechanisms that seem essential to our problematization. For example, in the case of a primary school, different attachment points may be identified, such as the relationship with the Ministry of Education. The researcher would then try to understand the organizational framework of this ministry, the digital tools it implements, its hierarchical structure, etc. The goal is to question the organization in the disciplinary areas that appear to present the strongest barriers in terms of detachment.

\subsubsection{Individual Interviews}

These interviews' purpose is to understand the situated representations and perceptions of actors within the organization, as well as to qualitatively identify their attachments to digital tools. It will be necessary to interview actors with different roles in the organization to cover a variety of perspectives, such as those from different hierarchical positions, localities, skills, etc. The qualitative research approach is not intended to obtain a statistical sample, but rather a theoretical one \cite{lejeune_manuel_2019}. Therefore, it is essential to question individuals who can provide viewpoints that will inform the researcher’s analysis. The interview grid may (and should) evolve during the survey. 

Our experience in exploratory fields has enabled us to identify several locks and levers to the ecological redirection of digital practices, presented in Table \ref{table_locks_levers}. Researchers may use these themes when preparing their own interview grids. However, we emphasize that this list is not exhaustive, and that the characterization of attachments is highly context-specific to the organization studied. Researchers should therefore quickly adapt their interview grids based on the first analytical elements from initial observations and early interviews.

\begin{table}[h!]
    \small
    \centering
    \caption{Suggestion of locks and levers for the redirection of digital practices to be addressed in the survey phase}
    \label{table_locks_levers}
    \begin{tabular}{>{\RaggedRight\arraybackslash}p{3cm} >{\RaggedRight\arraybackslash}p{4.8cm}}
        \toprule
        \textbf{Locks}&\textbf{Example of associated verbatim} \\
        \midrule
        Hierarchical obligation&\textit{"It is imposed by our organization, I have no choice"} \\
        \addlinespace
        Lack of knowledge and/or skills&\textit{"We don't have enough knowledge/skills to do otherwise"}\\
        \addlinespace
        Internal/external pressure (-)&\textit{"People expect/demand us to use it"}\\
        \addlinespace
        Lack of resources&\textit{"We don't have the time / means to change"}\\
        \addlinespace
        Habits&\textit{"We're used to it, and it's easier to stick with this way of doing"}\\
        \addlinespace
        Lack of alternative&\textit{"It's not possible to do without it"}\\
        \addlinespace
        Change unnecessary&\textit{"We don't have much reason to change" / "We hadn't even thought about changing"}\\
        \addlinespace
        Use appetence&\textit{"I like using it"}\\
        \addlinespace
        Direct gains&\textit{"It's actually super practical / efficient"}\\
        \addlinespace
        Fatalism&\textit{"The tool is developing anyway, might as well use it"}\\
        \addlinespace
        Enthusiasm for technology&\textit{"It broadens the range of possibilities"}\\
        \addlinespace
        \toprule
        \textbf{Levers} & \\
        \midrule
        Autonomy&\textit{"It isn’t imposed on us"}\\
        \addlinespace
        Sufficient skills and knowledge&\textit{"We know how to make the change"}\\
        \addlinespace
        Internal/external pressure (+)&\textit{"People expect/demand us to move away from it" / "People who don’t use it don’t inspire us"}\\
        \addlinespace
        Available resources&\textit{"We have the time and means to change"}\\
        \addlinespace
        Habits&\textit{"We're used to doing without it"}\\
        \addlinespace
        Existing alternative&\textit{We could do differently" / "We could do without it"}\\
        \addlinespace
        Change necessity&\textit{"I think it's important to do differently"}\\
        \addlinespace
        Aversion to technology&\textit{"I don't like this tool"}\\
        \addlinespace
        Direct gains&\textit{"It would save us time/money" / "It would be simpler without it" / "We would be more independent"}\\
        \bottomrule
    \end{tabular}
\end{table}

\subsubsection{Questionnaire}

Finally, we recommend that the researcher implement a questionnaire within the organization. This questionnaire serves two purposes: first, it provides a quantitative component for understanding the attachments to digital tools within the organization; second, it gathers data from a broader and more representative panel than those involved in the interviews and observations. This strengthens the confidence of the focus group (established for the following phases) and fosters participation from individuals with limited time to contribute to the survey. We suggest that the questionnaire last no more than 15 minutes and focus on quantifying, for selected digital tools, the function of the tool, the users, the perceived dependence of the organization on it, the individual dependence on it, and the main reasons for this (in)dependence.

\subsubsection{Investigation Report}

At the end of the investigation, the researcher provides an analysis report intended for the focus group, which will establish the action plan. This report enables the working group to anchor their work in the realities of the organization's legacy and to take into account these realities in their strategic redirection approach.

We recommend structuring the investigation report as follows:
\begin{enumerate}
    \item A summarized qualitative restitution of the survey in a free format. It may look like a classic qualitative sociological analysis, but in a condensed form. We suggest limiting it to no more than two pages, to make it accessible for the focus group.
    \item Attachment graphs associated to functions or activities carried out by the organization, where each point represents a digital tool. The vertical axis is defined by an environmental impact indicator, the horizontal axis by the degree of attachment, and a colour gradient reflects the usage scale of the tool. The environmental impact indicator is to be chosen by the protocol leader. The attachment degree and usage scale can be determined based on questionnaire responses. An example of such graph is provided in Figure \ref{figure_graph}. Other similar graphs can be found in different ecological redirection reports \cite{marchand_fermer_2023}.\footnote{Although life cycle assessment typically relates the environmental impacts of technical tools to the services they enable, this is not the case in this graph, which presents tools that fulfill different functions. This choice was made in order to present a single graph mapping all of the organization’s tools, so as not to overwhelm participants with too much information. Participants will therefore need to make the effort to isolate the tools serving the same function in order to carry out meaningful comparisons.}
    \begin{figure}[h!]
        \centering
        \includegraphics[width=0.5\textwidth]{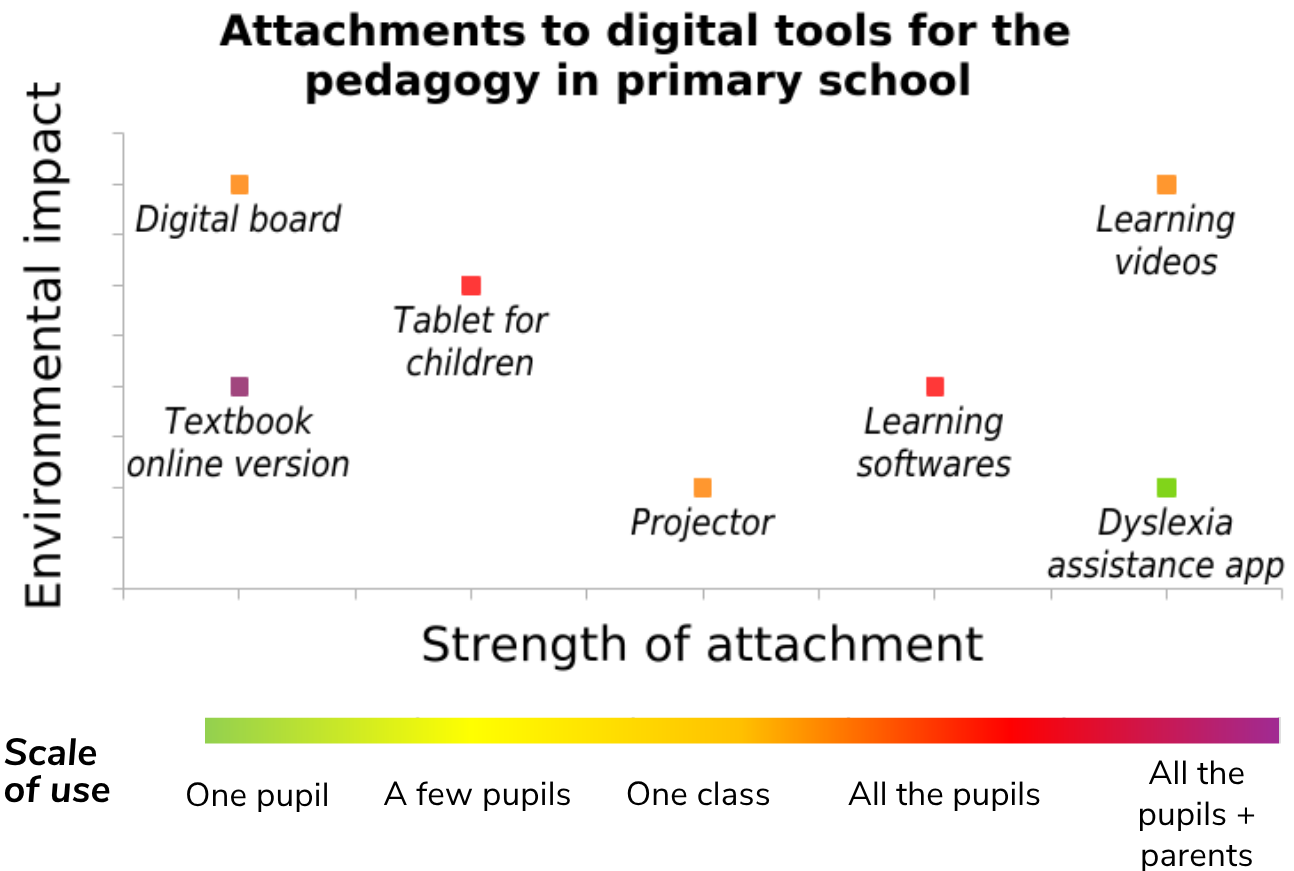}
        \caption{Example of an attachment graph}
        \label{figure_graph}
    \end{figure}
    \item For each digital tool, provide a fact sheet containing information such as: its functionality, scale of use, various impact indicators (data flow and storage associated, emissions, infrastructure required, quantity of minerals extracted in conflict regions, etc.)\footnote{This article does not provide a specific source for obtaining such information. If accessing this data is difficult, we recommend focusing on impact indicators that can be gathered through investigation and a basic understanding of digital infrastructures, for example data flows (in bytes per year), storage requirements of the tool (in bytes), or the type of infrastructure involved.}, an attachment indicator calculated from the questionnaire responses, and the main reasons for these attachments.
\end{enumerate}

\subsection{Phase 2 : Workshops}\label{sec_workshops}

Following this survey, working sessions are organized with the focus group to establish an action plan. These working sessions (workshops) will initially allow participants to understand the key issues about ICTs and their ecological redirection, in order to improve their knowledge on these topics.

Our experience with this approach has highlighted the challenges of negotiating the initial implementation of the protocol with an organization. This negotiation phase defines whether the organization is willing to engage in the conduction of the protocol or not and defines its modalities. It is clear that the time allocated per workshop and the number of participants are critical points in negotiations with organizations engaged in production imperatives. Therefore, to reach a compromise between negotiation and the quality of the workshops, we suggest organizing 2-3 hour sessions for this phase.

The first session is a presentation on the impacts related to ICTs and presents the investigation findings. We recommend including the following themes:

\begin{itemize}
    \item How the internet network functions
    \item Direct effects of ICTs
	\item Indirect effects of digitalization
	\item Holistic vision of the link between digitalization and acceleration
	\item Vulnerabilities of an over-digitized society
	\item Short-term individual benefits of reducing the digitalization of our lives
\end{itemize}

This session should aim to generate motivation among participants to engage in a transformation process \cite{lewin_field_1951}. This motivation should be encouraged through a sensitive (rather than purely statistical) presentation of the direct effects of ICTs, as well as by emphasizing the short-term positive effects of reducing digital practices among participants. We also suggest making this session open to the entire organization (and potentially the public) to raise awareness of these issues, create a broad movement of support for the protocol, and possibly recruit new members for the focus group.

The second session begins with a presentation of the ecological redirection concept and possible alternatives to digital technologies as they are currently deployed. It should outline both the technical (and non-technical) possibilities and impossibilities for redirection. This session then engages participants in a creative (writing) activity to envision desirable futures with less digital technology in their organization. This phase is important to stimulate creativity and avoid limiting the action plan to purely constraining considerations, which would restrict it to minor changes. This constraint mechanism was experimented with by the authors in workshops where personae with strong interests in maintaining the status quo were incorporated. The session concludes with the presentation of the format expected for the action plan, followed by a collective decision-making exercise to define the motivations and objectives that will guide the development of the proposed action plan.

At the end of the second phase, the protocol leader suggests participants to think individually about the measures to be included in the action plan before the next session. This long ideation period is essential to allow participants to think deeply about relevant actions to propose.

\subsection{Phase 3 : Debate on arbitrations}\label{sec_debate}

The objective of this phase, which is actually the third working session, is to enable negotiation around attachments to digital practices among the organization’s stakeholders. The workshops will facilitate an informed debate among participants regarding the necessary arbitrations and the formulation of an action plan. This action plan is not limited to incremental changes in the organization's digital practices: it reflects a strong positioning on the organization's strategic and political decisions, by formulating certain choices of renunciation and de-digitization of certain services. The notion of arbitration reflects this strategic decision-making dimension of the debate. This debate phase is crucial, and it must be guided to ensure that participants make full use of the resources from the investigation and workshops to support their arguments.

The session begins by establishing an environment of trust and respect within the group. The protocol leader reminds participants the session's objectives.

The first part of the action plan involves defining the motivations of the focus group for creating this action plan ("We believe it is important to implement this action plan because..."). Participants then name a follow-up reference person to ensure the proper implementation of redirection actions and the continuity of the protocol in the future, as well as a communication reference for communicating the plan within the organization. This accountability anchors the ecological redirection approach as a sustainable practice within the organization.

\begin{figure}[h!]
    \centering
    \includegraphics[width=0.5\textwidth]{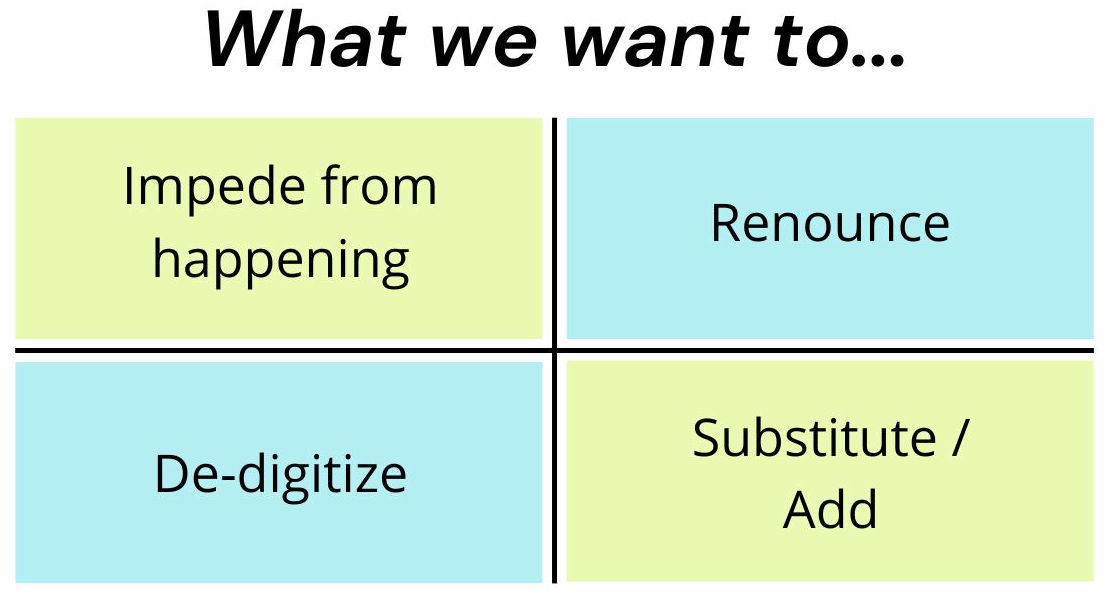}
    \caption{Categorization of actions for digital ecological redirection}
    \label{figure_action}
\end{figure}

Next, participants agree on a few key functions (ideally a small number to comply with the time constraints and to establish an achievable plan of action) currently handled by digital tools (as identified by the researcher), which they will debate. These could include internal communication, finances, human resource management, administration, creative activities, external communication, etc. For each function, participants suggest different digital sobriety measures. These measures should be understood in a broad sense, as they can take multiple forms: a political positioning of the organization, the replacement of a digital tool, the de-digitization of a practice, the introduction of a new business strategy, the discontinuation of a service, the restructuring of the organization, and so on. The range of possibilities is wide, and the potential impacts on the organization may vary in scale. A classification for these different categories of action is proposed in Figure \ref{figure_action}. Participant might use this supporting figure to expose their propositions by answer the question : "What we want to (as an organisation) ?"

The digital sobriety actions proposed by the focus group should be based on the material of the previous workshops. For each function explored, the group should:
\begin{enumerate} 
    \item Review the digital tools currently in use; 
    \item Identify the organization’s attachments to these tools; 
    \item Understand the environmental issues associated with these tools; 
    \item List potential alternatives to certain tools, if needed, in accordance with the vision of a desirable digital future developed during the imagination workshop; 
    \item Engage in debate to reach a consensual proposal. 
\end{enumerate}

Here are some clarifications regarding the structure of this deliberation process:

\begin{itemize}
    \item The facilitator plays a key role in guiding the focus group toward relevant propositions by supporting an informed and structured deliberation process.
    \item Digital tools with the greatest environmental impact should be prioritized when proposing sobriety measures. The level of ambition of these measures should be in accordance with the associated environmental impact.
    \item Attachments to digital tools must be taken into account when formulating proposals, using both quantitative and qualitative analysis. The stronger the attachment to a tool, the more carefully the proposed measures should be implemented so as not to disrupt the subsistence networks of those who rely on it, to ensure the availability of satisfactory alternatives\footnote{Proposing an alternative --whether digital or not-- is essential to the process of renunciation. This point of attention is emphasized by numerous academic authors working on sociological approaches to innovation through withdrawal \cite{goulet_faire_2022}.}, and to respect social justice. If we get back to our fictional example of a primary school, an application designed to support students with dyslexia may have a low environmental impact and be strongly valued by the students who use it. In such a case, renouncing the this tool would not be acceptable, as it would negatively affect those students. Instead, efforts should focus on tools with greater environmental impact.
    \item The objective of these discussions is to reach consensus within the focus group for each proposal. If consensus cannot be achieved, a vote may be used to resolve non-consensual decisions.
\end{itemize}

Finally, the protocol leader offers participants, especially the follow-up reference person, resources to maintain a redirectionist posture (questioning the need and compatibility of practices with planetary boundaries) within the organization.

\subsection{Phase 4 : Implementation of closures}\label{sec_implementation}
The decision phase involves the implementation of the chosen redirection measures. These measures will necessarily involve renunciations of certain practices and, at a minimum, changes to others. It is necessary to support these closures and transformations both technically (especially in terms of digital infrastructure) and humanely (to reorganize the organization’s activities, according to the new practices). This article does not propose a generic solution for this implementation phase, given the highly specific nature of each situation encountered. However, it emphasizes the importance of engaging technically and organizationally competent individuals to carry out the anticipated closures. Optional exchange sessions for organization members may be arranged during this phase.

\section{Case Studies and Expected Results}

This protocol for the ecological redirection of digital tools and practices is currently tested through two case studies to identify the challenges of the practical implementation and to improve both the protocol itself and the supporting documentation for its application. 

\subsection{Case studies}
Two case studies will be conducted: the first in an eco-village and the second in a town hall. This experimental work has already begun and has led to the evolution of the protocol --particularly the survey resources-- toward the version presented in this article. The two case studies present different characteristics, and the experimentation is at varying stages of advancement. The main characteristics of these two experiments are presented in Table \ref{table_case_studies}.

\begin{table}[h!]
    \small
    \centering
    \caption{Characteristics and stage of ongoing experiments in the ecological redirection protocol}
    \label{table_case_studies}
    \begin{tabular}{p{2.4cm}|p{2.5cm}|p{2.5cm}}
        \toprule
        \textbf{Case Study} & Eco-village & Town Hall \\
        \addlinespace
        \textbf{Legal structure} & Cooperative company of collective interest & Public legal entity (territorial collectivity) \\
        \addlinespace
        \textbf{Focus group} & Voluntary members & Voluntary employees \\
        \addlinespace
        \textbf{Intensity of digital use} & Low & High \\
        \addlinespace
        \textbf{Progress of the experimentation} & Phase 1 ongoing: Analysis of survey material & Phase 1 ongoing: Data collection \\
        \bottomrule
    \end{tabular}
\end{table}

\begin{table*}[h!]
    \small
    \centering
    \renewcommand{\arraystretch}{1.3}
    \setlength{\tabcolsep}{5pt}
    \begin{tabular}{>{\raggedright\arraybackslash}p{2cm}         >{\raggedright\arraybackslash}p{4cm} >{\raggedright\arraybackslash}p{3.5cm}     >{\raggedright\arraybackslash}p{3cm} >{\raggedright\arraybackslash}p{4cm}}
        \toprule
        \textbf{Phase} & \textbf{Steps} & \textbf{Resources given}\footnotemark[7] & \textbf{Required Skills and Knowledge} & \textbf{Expected Deliverable} \\
        \midrule

        Mapping of attachments
        \newline (section \ref{sec_mapping})&
        1. Observation \newline
        2. Questionnaire \newline
        3. Interviews &
        - Questionnaire Template \newline
        - Interview grid template \newline
        - Investigation report template &
        - Human and social sciences method knowledge &
        A report including: \newline
        - A summarized qualitative restitution of the survey \newline
        - Attachment graphs \newline
        - Fact sheet for each digital tool \\

        \addlinespace

        Workshops
        \newline (section \ref{sec_workshops}) &
        1. Workshop on digital sector impacts and vulnerabilities + Restitution of the mapping of attachments phase \newline
        2. Workshop on alternative digital practices + imagination on desirable futures + action plan presentation + reflection exercise on motivations and objectives  &
        - Presentation material on digitalization's impacts and vulnerabilities + alternative digital practices \newline
        - Roadmap to conduct the different workshops \newline
        - Imagination exercise&
        - Knowledge on digital sector impacts and vulnerabilities + alternative digital practices \newline
        - Skills in workshop animation &
        Story created by the participants during the imagination workshop \\

        \addlinespace
        
        Debate on arbitrations
        \newline (section \ref{sec_debate})&
        1. Set up of an environment of trust and respect \newline
        2. Definition of the terms of the action plan \newline
        3. By function, arbitration and proposal of actions to be implemented &
        - Template of action plan for ecological redirection of the organization \newline
        - Debate process reminder&
        - Skills in debate animation &
        Completed action plan for ecological redirection of the organization \\
        
        \addlinespace

        Implementation of closures 
        \newline (section \ref{sec_implementation}) &
         &
         &
        - IT skills \newline
        - Skills in other different domains (economy, law, strategy, planning, social skills, …) &
        Report documenting achievements, failures, and causes of these failures \\
        \bottomrule
        \end{tabular}
    \caption{Recap table of the ecological redirection protocol for digital practices}
    \label{table_summarize}
\end{table*}

\subsection{Results Analysis}
The analysis of the results from this experimentation will focus on three main objectives: (1) the fluidity of the protocol’s execution, (2) its effectiveness in redirecting digital practices, and (3) the identification of socio-technical barriers that hinder redirection.

\subsubsection{Fluidity}
Experimenting this protocol on case studies will enable us to raise difficulties relating to its form. These obstacles will be measured by a post-experimental questionnaire in which participants will be able to talk about their failures, their perceptions, the difficulties they encountered, etc. Recommendations may be put forward by the participants themselves and will enable the modalities of the different phases to evolve.

\subsubsection{Effectiveness}
The effectiveness of the ecological redirection protocol can be evaluated on the two case studies, thus highlighting the degree of relevance of its implementation on a wider scale. Indicators will be proposed to quantify this redirection, based in particular on residual digital materiality in the organization (number of terminals, digital services, screen time, etc.), as well as on digital attachments, quantified by a post-experimental questionnaire. This quantification will be completed by a narrative creation exercise based on participants' testimonials, in order to illustrate the benefits of the protocol.

\subsubsection{Socio-technical Barriers}
The renunciation of digital tools is hindered by socio-technical barriers of various kinds and at different scales. Ratchet effects are systemic phenomena that prevent a reversal in the development of technology within society, and digital technologies are significantly affected by these effects \cite{girard_computing_2024}. At the individual level, various psychological processes make detachment from digital technologies difficult \cite{goncalves_digital_2023}. However, these mechanisms do not adequately explain collective dependencies on digital technologies, as collective dependency does not simply equate to the sum of individual dependencies \cite{noauthor_when_2023}. Therefore, a qualitative inquiry will be conducted following the implementation of the case studies to properly objectify these socio-technical barriers at the collective level. The differing nature of the two experimental settings will allow for the identification of contextual factors within organizations that influence these collective lock-ins.

\subsection{Challenges and Discussions}

Only the initial stages of this protocol have been implemented in the two case studies (see Table \ref{table_case_studies} for the current state of these experiments). These early experiments have helped to identify several challenges to be addressed for the concrete implementation of the protocol. This section outlines some of these challenges and opens a discussion on its relevance.

\subsubsection{Uncertainty in Mapping Attachments}

The investigation phase raises questions about the very possibility of mapping attachments, as actors are not always aware of their own. The notion of attachment is central to the academic field of ecological redirection, yet its exact definition has not been fully formalized in the literature. Attachments are studied in diverse forms and through various disciplinary fields : philosophical, sociological, systemic, and others.

Sociological literature suggests that attachments are truly revealed only when they are threatened or at least challenged \cite{Hennion_attachements_2010}. For instance, it is often only during a shortage of a common consumer good that users become aware of its importance in their daily lives.

Accordingly, when conducting an investigation for the mapping of attachments, the investigator can only access the actors’ perceived attachments at the time of the investigation. An external event, such as a disruption to digital tools or the implementation of digital sobriety policies within the organization, may reveal attachments that were previously invisible.

For these reasons, the inquiry should not be considered as limited to this initial mapping stage. Throughout the entire duration of the protocol, and even beyond --during internal or external organizational changes-- it is important to remain attentive to how attachments evolve and manifest.

Ecological redirection through investigation, while formalized in this article as a structured protocol, should also be understood as a continuous and iterative process. In this process, investigation serves as an ongoing exploration of the issues surrounding actors’ attachments and the design of new ways of inhabiting a world disrupted by socio-environmental collapses.

\subsubsection{Involvement of Organizations}

During the search for case studies, the organizations that agreed to test the protocol expressed strong constraints regarding the time available and their capacity to invest their members in this experiment. Organizations engaged in activities unrelated to digital sobriety, or whose daily routines are saturated with productive tasks, are often only able to dedicate a few hours (and only by a small number of members) to this initiative.

Yet, the experience of ecological redirection is a long-term process. It requires question deep certainties about how our society and the Earth system functioning. It involves a transformation in how the world is understood, and a deconstruction/reconstruction of the futures the organization had projected for itself. These stages demand time, sustained support, and must move through the lived experience of the organization's members. This disruption of certainties leads actors into a state of disorientation or "trouble" \cite{cuny_redirection_2025}, which becomes the driving force behind a radical reorientation of the organization’s trajectory.

Therefore, given the relatively limited investment required from the organization in light of the magnitude of the issues at stake, the current version of the protocol is unlikely to achieve a full-scale redirection of the participating organization.

However, it does succeed in overcoming pragmatic barriers of acceptability. Indeed, an organization would be unlikely to commit to a demanding, time-intensive process without first having encountered the urgency of digital sobriety in concrete terms. The protocol thus acts as an entry point (a foot in the door) allowing the organization to begin to experience the real issues of the Anthropocene and to sense the urgency for redirection. For this reason, implementing a follow-up strategy to sustain the redirection process after the protocol’s completion is essential. In this way, the investigation process --which is central to ecological redirection-- must go beyond a top-down approach from an investigator to a group of respondents: it is the investigated actors themselves who must carry out the investigation, in order to collectively understand and define the trouble they are experiencing. The protocol should be seen as the initiation of a long-term, non-linear, and iterative process of ecological redirection, one that must permeate the organization, affect its members, and create trouble in projections toward futures that are incompatible with planetary boundaries.

\section{Conclusion}
Ecological redirection is a concept that seeks to propose new ways of approaching sustainability, focusing on the question of direction rather than technical means, with a strong intention to remain rooted in the heritage of the techno-industrial society and to introduce renunciations and closures within it. This theoretical framework remains highly conceptual and requires operationalization for organizations. This paper proposes a protocol for the ecological redirection of digital tools and practices within organizations, based on four phases: mapping of attachments, workshops, debate on arbitrations, and implementation of closures. This protocol aims to support organizations in taking radical and concrete action toward digital de-escalation. It offers a grip for individuals who feel powerless regarding the digitization of society and the rising impacts related to it.

While this explicit protocol facilitates the implementation of ecological redirection in organizational activities, it is important to understand that this practice is strongly dependent on the political context in which organizations operate. The issue of renunciation and closure of certain industrial activities are currently difficult to even consider for most actors, particularly industrial ones, because of the neoliberal capitalist political framework in which they are. It is crucial to emphasize that the protocol for the ecological redirection of digital practices proposed in this article will be difficult to generalize within a techno-industrial society not engage in strong environmental policies and still settled in a regime of economic growth \cite{parrique_political_2019}. 

Therefore, this protocol should not be seen as a solution to a socio-environmental crisis, but rather as an operational support for ambitious policies aimed at reorganizing the conditions of subsistence in society during the Anthropocene. Ecological redirection protocols should not be limited to isolated projects aiming only to altering specific practices and infrastructures. These protocols should have transformative aims in a radical sense \cite{gersick_revolutionary_1991} for organizations conducting them. They must initiate a real momentum of awareness, politicization, and action in response to the Anthropocene. Ecological redirection seeks to trigger this transformation through survey and design, within an approach of heritage, arbitration, and closure, but also aims to initiate a broader dynamic of politicization of the techno-industrial model, to make it land towards a desirable and sustainable society. Organizations shall be aware of this transformative aim while conducting such protocols, in order to formulate radical change policies for their ecological redirection.

\bibliographystyle{ACM-Reference-Format}
\bibliography{references.bib}

\appendix

\end{document}